\documentstyle[aps,prl,twocolumn,epsfig]{revtex}

\newcommand{\ob}{\bar{\omega}}

\begin{document}

\twocolumn[\hsize\textwidth\columnwidth\hsize\csname@twocolumnfalse\endcsname

\title{Bose-Einstein Condensation of Metastable Helium}
\author{F. Pereira Dos Santos,  J. L\'eonard,  Junmin Wang$^{a}$,  C. J. Barrelet,\\
F. Perales$^b$,  E. Rasel$^{c}$,  C. S. Unnikrishnan$^{d}$,  M. Leduc,  C. Cohen-Tannoudji\\
Coll\`ege de France,\\
Laboratoire Kastler Brossel$^*$, D\'epartement de Physique, Ecole Normale Sup\'erieure, \\
24 rue Lhomond, 75231 Paris Cedex 05, France}

\date{\today}    
\maketitle

\begin{abstract}
We have observed a Bose-Einstein condensate in a dilute gas of $^4$He in the $2^3S_1$ metastable state. We find a critical temperature of $(4.7\pm0.5)\,\mu$K and a typical number of atoms at the threshold of $8\times10^6$. The maximum number of atoms in our condensate is about $5\times10^5$. An approximate value for the scattering length $a=(16\pm8)$ nm is measured. The mean elastic collision rate is then estimated to be about $2\times10^4$ s$^{-1}$, indicating that we are deeply in the hydrodynamic regime. The typical decay time of the condensate is $2$ s, which places an upper bound on the rate constants for 2-body and 3-body inelastic collisions. 

\


\

\end{abstract}]

Bose-Einstein condensation (BEC) of dilute atomic gases was first observed in alkali
atoms in 1995 and then, a few years later in atomic hydrogen. Since then, the field
has developed in a spectacular way both experimentally and theoretically \cite{School99}. 
So far only condensates with atoms in their electronic ground state have been produced. 

Several laboratories are currently involved in the search for BEC of atoms 
in an excited state, namely noble gases in an excited metastable state. Helium in its triplet metastable $2^3S_1$ state ($^4$He$^*$) is of particular interest. The first advantage of $^4$He$^*$ is its large internal energy (19.8 eV). It allows for a very efficient detection 
of the atom by ionization after collision with another atom or a surface, which can be of interest for atomic lithography \cite{Bard97,Nowak96}. Second, helium is a relatively simple atom which allows for quasi-exact calculations that are useful in metrological applications.
Third, mixtures of $^3$He and $^4$He can be used to study quantum degenerate
mixtures of bosons and fermions. Finally, Penning collisions are expected to be inhibited for spin polarized atoms due to spin selection rules. This effect, first pointed out in \cite{Shlyap94}, was confirmed by subsequent calculations \cite{Shlyap96}.

The present article describes the observation of BEC of $^4$He$^*$ atoms. Similar results have also been obtained at IOTA, Orsay \cite{BecOrsay}. The two experiments differ by their detection methods. The Orsay group detects the atoms falling on a microchannel plate, whereas we use an optical absorption imaging of the atomic cloud on a CCD camera. The two experiments therefore give different and complementary information on the physics of BEC in $^4$He$^*$.

The first step of our experiment is the efficient loading of a magneto-optical trap (MOT). 
The experimental set-up is described in detail in \cite{Per101}. 
A discharge atomic source ensures a high flux of triplet metastable atoms of $10^{14}$ atoms/s.sr, with a mean velocity of about 1000 m/s. The atomic beam is collimated \cite{Rasel99} and Zeeman slowed by laser light at 1083 nm ($2^3S_1$-$2^3P_2$ transition). A narrow frequency band master oscillator (DBR diode laser) injects
a Yb-doped fiber amplifier with an output power of 500 mW. Using this set-up,
it is possible to trap $\sim8\times10^8$ atoms in the MOT at a temperature of 1 mK.

\begin{figure}[htb]
\begin{center}
\epsfig{file=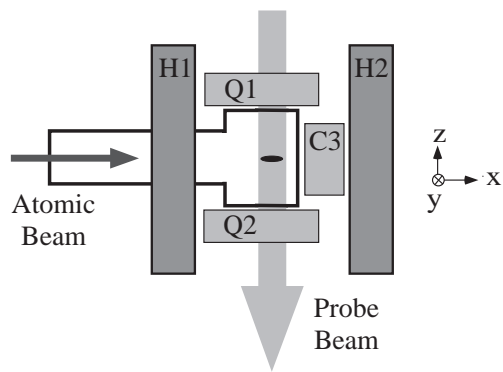,height=5cm,width=7cm}
\end{center}
\caption{\footnotesize Top view of the magnetic trap. Coils $Q_1$ and $Q_2$ produce a quadrupole field used for the MOT. Combined with a third coil $C_3$, they produce a magnetic Ioffe-
Pritchard trap. Helmholtz coils $H_1$ and $H_2$ compensate the 
bias field. The probe beam used for optical detection is perpendicular to the longitudinal 
axis of the trap.}
\label{piege}
\end{figure}

$^4$He$^*$ atoms are confined at the center of a small ($4\times4\times5$ cm) quartz cell. All coils are external to the cell (see Fig.\ref{piege}). The coils $Q_1$ and $Q_2$ (144 turns and 7 cm diameter) combined with the coil $C_3$ (108 turns and 4 cm diameter) produce an anisotropic magnetic Ioffe-Pritchard trap. Two additional Helmholtz coils reduce the bias field, in order to increase the radial confinement of the trap. A current of 45 A in all the coils produces a 4.2 G bias field, radial gradients of 280 G/cm and an axial curvature of 200 G/cm$^2$. These values correspond to trapping frequencies of 115 Hz in the axial and 1090 Hz in the radial directions. The current in the coils can be switched off in 200 $\mu$s.

The second step of the experiment is the loading of the magnetic trap. After switching off 
the MOT field, the cloud is further cooled down to about 300 $\mu$K during a 1 ms optical molasses phase. To increase the transfer efficiency from the molasses to the magnetic trap, the atoms are optically pumped by a circularly polarized laser pulse. $3\times10^8$ atoms are loaded in the magnetic trap. The lifetime of the atomic cloud in the magnetic trap is about 35 seconds, and its temperature is 1.2 mK after compression and bias compensation. 

The last step of the experiment consists of evaporative cooling performed by radio frequency 
(RF) induced spin flips. The frequency is ramped down from 160 MHz to around 12 MHz in 15 seconds. After evaporation, the trap is switched off and the cloud released from the trap is probed by absorption imaging on a CCD camera whose quantum efficiency is 1.5\% at 1083 nm.

\begin{figure}[htb]
\begin{center}
\epsfig{file=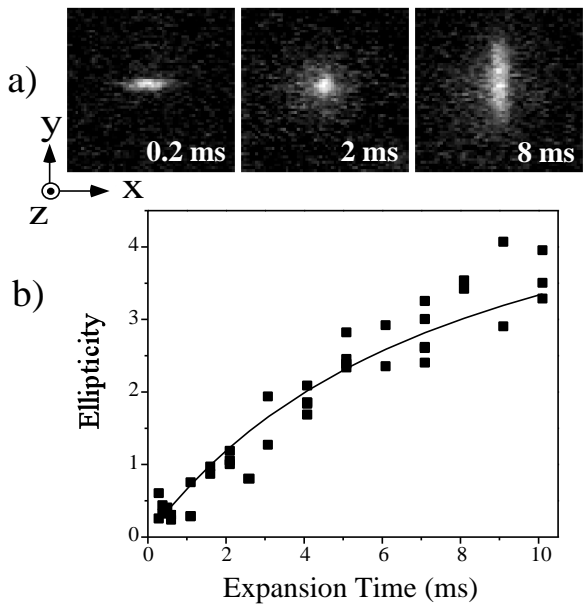,height=8cm,width=8cm}
\end{center}
\caption{\footnotesize a) Time-of-flight absorption images ($1.4\times1.4$ mm) after an expansion time of 0.2, 2 and 8 ms; b) Ellipticity of the condensate for increasing expansion times. The solid line is the theoretical prediction without any adjustable parameters. The inversion of ellipticity is characteristic of the behaviour of an expanding condensate.}
\label{tof}
\end{figure}

When the RF frequency of the evaporation is ramped down to a final frequency below 13 MHz, a narrow structure appears on the absorption image which we identify as a condensate. The strongest evidence for the presence of the condensate is the evolution of the shape of
this structure when released from the trap in the time-of-flight (TOF) measurement. Its anisotropy increases as the expansion time is increased, and its ellipticity undergoes an inversion (see Fig.\ref{tof} a). This observation is a consequence of the mean-field interaction between atoms in the condensate \cite{Yvan}. The theoretical prediction containing no adjustable parameters agrees well with our measurements (see Fig.\ref{tof} b).

\begin{figure}[htb]
\begin{center}
\epsfig{file=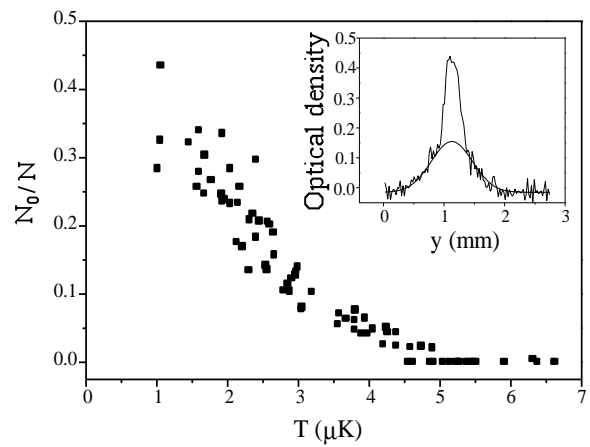,height=6cm,width=8cm}
\end{center}
\caption{\footnotesize Condensed fraction $N_0/N$ versus $T$. The insert shows a 1D profile of an absorption image, displaying a bimodal structure, composed of a condensate and a thermal component, which is characteristic of the bosonic gas below $T_c$. The thermal component was fitted for clarity.}
\label{Tc}
\end{figure}

The spatial distribution of the absorption pictures is fitted with the sum of two functions, one 
for the condensate and one for the thermal cloud (see insert in fig. \ref{Tc}). The function for the condensate
is an integration along the z-axis of a paraboloidal distribution. It describes the equilibrium density profile of the condensate within
the harmonic trap in the Thomas-Fermi limit \cite{String99}. The function for the thermal cloud is a $g_2$ 
function valid for a bosonic gas close to the transition where the chemical potential
$\mu$ is an adjustable parameter \cite{School99}. From the fit, we extract the ratio
between the number of atoms in the condensate, $N_0$, and the total number of atoms, $N$. 
Plotting $N_0/N$ versus the temperature $T$ (Fig.\ref{Tc}) gives the value of the critical temperature $T_c=4.7\pm0.5\,\mu$K, which is confirmed by a TOF measurement performed 
on a thermal cloud just above the transition. 

The results presented until now are based only on measured sizes of the clouds and on relative numbers like $N_0/N$ which do not require absolute calibration of the optical detection. The procedure described previously for the measurement of ellipticity and critical temperature is not suitable for absolute measures of $N$ and $N_0$. So, we use a different procedure to calibrate the number of atoms at threshold. The current is switched off in two steps: first, in the Helmholtz coils, then, after a delay of 10 ms, in the 3 coils $Q_1$, $Q_2$ and $C_3$, whereas the current in all five coils was switched off simultaneously for the measurements above. Then, we take an absorption image with an exposure time $T_{exp}=200\,\mu$s after a TOF time of 5 ms. This procedure minimizes the effects of eddy currents of the Helmholtz coils which last several milliseconds, Zeeman shifting the atomic resonance with respect to the probe beam frequency \cite{noteEC}. 
There are two other obvious sources of sensitivity losses. First, the width of the absorption lineshape is measured to be twice the natural linewidth, which we attribute to laser linewidth and power broadening. This reduces the absorption cross-section by a factor 2. Second, the probe beam propagates perpendicularly to the x-axis of the Ioffe-Pritchard trap (see Fig.\ref{piege}) along which atoms are polarized. It would therefore be necessary to take into account the populations of the different Zeeman sublevels, and the Clebsch-Gordan coefficients of the various Zeeman optical transitions excited by the probe beam. But any residual magnetic field redistributes the populations among Zeeman sublevels. Assuming equally populated sublevels gives an extra loss by a factor 9/5. Finally, there is another source of losses which is specific to $^4$He$^*$ when optically detected and particularly important at the high densities obtained in our experiment. 
The metastable atoms have a huge Penning ionization cross-section in the presence of resonant light, so that losses can accumulate during the probe pulse at large densities. Unfortunately, probing at low intensity is no longer possible if $T_{exp}$ is made shorter because of the low efficiency of the CCD camera. This long exposure time also increases the acceleration of the atoms due to radiation pressure which pushes them out of resonance \cite{noteSW}. In the present stage of the experiment, it is difficult to give a quantitative description of the combined effect of these phenomena. So, we prefer to wait an expansion time of 5 ms to sufficiently reduce the atomic density. Indeed, we observe that the measured number of atoms is an increasing function of the expansion time reaching a plateau after 5-6 ms. After the corrections mentioned above, the total number of atoms $N_c$ at the threshold is measured to be 5$\times10^6$ atoms with an accuracy of about 50\%.

$N_c$ can also be estimated from $N_c=1.202(k_BT_c/\hbar \ob)^3$, where $\ob$ is the geometrical average of the frequencies of the trap \cite{deviation}. We deduce $N_c=8.2\times10^6$ atoms with an uncertainty of about 30\% compatible with the previous value. Because it is easier to trace down the error on $N_c$ when derived from $T_c$, we arbitrarly choose this value of $N_c$ to estimate the scattering length $a$. Also, assuming that the transition occurs at a phase space density equal to $n(0)\lambda_{dB}^3=2.612$, where $\lambda_{dB}$ is the de Broglie wavelength, we can also derive the density at the center of the trap $n(0)=(3.8\pm0.7)\times10^{13}$ atoms/cm$^3$ at the transition.

Within the Thomas-Fermi approximation, one can extract the chemical potential $\mu$ from 
the size of the condensate \cite{String99}. As the optical detection around the transition 
has been calibrated, we can now use our previous relative measurements to deduce the absolute number of condensed atoms $N_0$ below the transition. Typical values of $\mu=1.4\times10^{-29}$J and $N_0=(4\pm1.5)\times10^{5}$ are obtained with condensates prepared at temperatures ranging from 1.2 $\mu$K to 3 $\mu$K. Finally an estimation of the scattering length $a$ can be given using $a=\sigma/15N_{0}\times(2\mu/\hbar\ob)^{5/2}$, where $\sigma =(\hbar/m\ob)^{1/2}$ is the characteristic size of the ground state of the trap. We find $a=(16\pm8)$ nm which is compatible with the value given by recent theoretical works \cite{Shlyap96,Vent00}. The error is mostly due to the uncertainty on the number of atoms. Knowing the scattering length $a$, the density at the center of the trap $n(0)$ and the critical temperature $T_c$, we obtain a mean rate of elastic collisions $\bar{\gamma}_{coll}\simeq2\times10^4$ s$^{-1}$ near threshold, leading to $\ob/\bar{\gamma}_{coll}=0.17$. We thus enter in the hydrodynamic regime, an interesting feature for a gas above $T_c$ \cite{Ketterle,Griffin,DavidGO,Wu98}.
 
\begin{figure}[htb]
\begin{center}
\epsfig{file=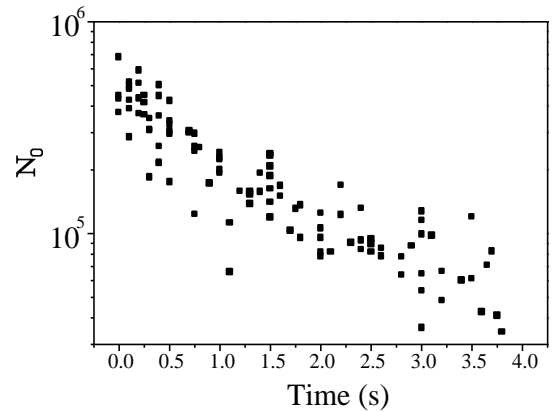,height=5.6cm,width=7.5cm}
\end{center}
\caption{\footnotesize Decay of the condensate. This measurement was performed 
with an RF shield at 12.3 MHz.}
\label{lifetime}
\end{figure}

Fig.\ref{lifetime} shows the evolution of the number of atoms in the condensate versus the trapping time. The typical lifetime is about 2 s. In the present stage of the experiment, it is not possible to discriminate between 2-body or 3-body decay. However, assuming that only 2-body collisions lead to losses in the condensate, we can place an upper bound on the collision rate constant $G$ between spin polarized $^4$He$^*$ atoms, which is predicted to be inhibited by a factor 10$^4$ compared with the rate constant for Penning ionization collisions between unpolarized atoms. The evolution equation of the number of atoms $\dot{N}(t)=-G\int n^2(\bf r\rm)d^3r$, integrated on the whole condensate in the Thomas-Fermi approximation, gives the evolution of the number of atoms $N_0$ in the condensate. A fit leads to $G\leq(4.2\pm0.6)\times10^{-14}$ cm$^3$/s, which corresponds to a reduction factor larger than $2\times10^{3}$, much larger than the previously measured ones \cite{Tol00,Nowak00}, in agreement with theoretical calculations \cite{Shlyap94,Shlyap96}.

If one assumes now that 3-body collisions are responsible for the decay of $N_0$, one can 
as well give an upper bound for the rate constant $L$ defined by $\dot{N}(t)=-L\int n^3(\bf r\rm)d\rm^3r$. Fitting our data gives $L\leq(2.8\pm0.2)\times10^{-27}$ cm$^6$/s. This value can be compared to theoretical predictions \cite{Shlyap3,Esry99}: for example, our upper limit is compatible with \cite{Shlyap3}, which finds $L=3.9\hbar a^4/2m=2\times10^{-27}$ cm$^6$/s. 

In conclusion, this article shows the evidence for the formation of a BEC of helium atoms in the metastable state $2^3S_1$. Our results concerning the losses due to inelastic collisions show that the spin polarization does inhibit the Penning ionization collisions between 2 metastable helium atoms by more than 3 orders of magnitude, as theoretically predicted.
We find a large scattering length, which results in very large rates of elastic collisions : the cold gas at threshold is in the hydrodynamic regime, which we plan to study in more details. We also plan to further exploit the original characteristics of this new born condensate of atoms in an excited state.

\

\noindent {\bf Acknowledgments:}
The authors wish to thank J. Dalibard, C. Salomon, D. Gu\'ery-Odelin and Y. Castin for helpful discussions and careful reading of the manuscript, and F. Pavone and A. Sinatra-Castin for their contribution in the early stages of the experiment. We thank the IOTA goup in Orsay for communicating their preliminary results as soon as they were obtained, which was a great stimulation for our group achieving helium BEC 8 days later. 
Also, we thank M. Roux for the loan of the Hamamatsu CCD camera (C4880-30) with which the BEC was first detected in our group. C.J.B. acknowledges the support from the Schweizerische Studienstiftung.

\

$^a$ Permanent address: Institute of Opto-Electronics, Shanxi University, 36 Wucheng Road, Taiyuan, Shanxi 030006, China.

$^b$ Permanent address: Laboratoire de Physique des Lasers, UMR 7538 du CNRS, Universit\'e Paris Nord, Avenue J.B. Cl\'ement, 93430 Villetaneuse, France.

$^c$ Present address : Universit$\ddot{\rm{a}}$t Hannover, Welfengarten 1, D-30167 Hannover, Germany.

$^d$ Permanent address : TIFR, Homi Bhabha Road, Mumbai 400005, India.

$^*$ Unit\'e de Recherche de l'Ecole Normale Sup\'erieure et de
l'Universit\'e Pierre et Marie Curie, associ\'ee au CNRS (UMR 8552).


\begin{thebibliography}{30}
\bibitem{noteEC} Because of the large size of the Helmholtz coils, one can however neglect the spatial gradients of magnetic fields produced by eddy currents. The measurement of the sizes and of the ratios $N_0/N$ is thus not affected by these remaining magnetic fields.
\bibitem{noteSW} Actually, we use a standing wave made of two independent opposite beams
in order to reduce the effect of radiation pressure.
\bibitem{deviation} This formula is only valid in the case of an ideal bosonic gas
and does not require any further assumption. Effects due to mean field have been calculated 
for an harmonic trap \cite{String96} and effects due to quantum correlations have been calculated in a box \cite{Baym99}. The shift of the critical temperature $T_c$ can be quite large ($\delta T_c/T_c \simeq$ 15\%) for both of these effects, due
to the large value of the scattering length $a$, but have opposite signs. It would be interesting to study them both in more detail for an harmonic trap.
\end{thebibliography}
\end{document}